# Behavioural Predictors that Influence Digital Legacy Management Intentions among Individuals in South Africa

Jordan Young[1][0000-0000-0000-0000], Ayanda Pekane[1][0000-0002-4756-6357] and Popyeni Kautondokwa[1][0000-0001-9001-7313]

[1] University of Cape Town, Cape Town, 7700, South Africa

**Abstract.** An emerging phenomenon, digital legacy management explores the management of digital data individuals accumulate throughout their lifetime. With the integration of digital systems and data into people's daily lives, it becomes crucial to understand the intricacies of managing data to eventually become one's digital legacy. This can be understood by investigating the significance of behavioural predictors in shaping digital legacy management. The objective of this study is to explore how behavioural predictors influence the intentions of individuals in South Africa towards managing their digital legacy. This entailed: 1) investigating the impact of attitude, subjective norms, and perceived behavioural control on these intentions; 2) exploring the perceived usefulness of digital legacy management systems; and lastly 3) understanding the implications of response cost and task-technology fit on individuals' inclinations towards digital legacy planning. Data were collected (n = 203 valid responses) from South African residents using an online survey and analysed using partial least squares structural equation analysis (PLS-SEM). Results indicate that attitudes, peer opinions, personal resources and skills are significant positive influences on digital legacy management intention. Recognizing and understanding these behavioural predictors is key when developing region-specific and culturally sensitive digital legacy management tools, awareness campaigns and policies. Furthermore, it could pave the way for more tailored strategies, ensuring effective transfer of post-mortem data, reducing potential conflicts, and providing clarity when dealing with post-mortem data.

## 1    Introduction

During the 21st century, the value of digital objects has drastically increased becoming what we now refer to as digital assets [13] . Historically, legacies typically comprised of tangible items that were passed down from one individual to another, yet these physical items have been largely replaced by digital equivalents [47]. Digital data often serves as an extension of oneself and can be of high importance to individuals [12]. As people spend more time in their virtual lives, it is important to protect what we leave behind [7]. However, the shift towards digital versions of physical items has not been followed by a corresponding shift in the way these digital items are managed and passed down to future generations [47].

Digital legacy refers to the accumulated digital information, across various platforms and formats, left behind by an individual after their death, which includes content they created, shared, or that pertains to them [12, 13, 47]. Very few people have planned for their digital legacy as it is a relatively new concept [18]. The concept of digital legacy has recently been brought to light through social media and digital platforms such as Facebook [13] where guidelines for deceased user accounts are still emerging [9]. Managing one's digital legacy and understanding these behavioural predictors impacts: 1) the bereaved who may seek access or closure; 2) the way digital platforms handle one's data after one's death; and 3) the deceased's privacy, as their personal content could be accessed in undesired ways.

Due to the increase in data stored online and how frequently users interact online, there is growing concern about how valuable digital assets can be managed and passed on to loved ones [31]. Even death-aware people have not considered their own digital legacy [44]. The consequences of not managing one's digital legacy could include copyright infringement, invasion of privacy, theft of identity, and loss of private data [29]. Recognizing behavioural predictors is key to understanding how individuals approach digital legacies and tailored digital legacy management strategies can be developed, reducing conflicts and uncertainties. The implications of unmanaged digital legacies range from emotional distress of the bereaved to potential legal issues.

The objective of this study is to understand the behavioural predictors that influence digital legacy management intentions among individuals in South Africa. Understanding behavioural predictors is vital for the future of effective digital legacy management. As South Africa's digital landscape rapidly expands, it becomes imperative to address the complexities of managing and safeguarding digital assets for future generations [3, 6].

## 2 Literature Review

### 2.1 Digital Legacy Platforms

In recent years there has been a rise in digital afterlife research and many digital legacy platforms have been developed such as Afternote [30, 32]. Afternote is a digital platform that enables users to save their personal history, leave messages for loved ones, and record their final wishes [29]. These digital legacy platforms, designed for death-related practices that aim to preserve the memories of loved ones currently have small user bases; however, mainstream social media platforms have been gaining popularity amongst users for grieving and mourning-related posts [12]. Many startups have been capitalizing and thriving in the end-of-life sector, which is expected to continue as technology infiltrates society [29]. The digital afterlife industry ranges from small business applications like Afternote to larger ones like Facebook. Some digital afterlife platforms are free, and others require a fee, but these fees are based on a rate not intended for the standard of living in the Global South [29].

It is important not to blindly adopt unsuitable solutions from Western society, which refers to North American and European regions with predominantly Eurocentric values and perspectives. Despite this, popular platforms in the Global South, specifically in the context of South Africa are currently based on Western-influenced policies and guidelines.

### 2.2 Social Networking Sites

With the increase in Social Networking Sites (SNS) and trends suggesting that the rise in technology usage will continue, it is crucial to understand the long-term consequences for users in the context of end-of-life [8]. Some view the interaction and information on SNS as one's 'digital soul' that will become one's digital legacy [8]. Previous research has been done on the social relationships that form between the bereaved and deceased through social media [34]and how SNS manage their responsibility as a digital memorial site [33]. These SNS create a space for the bereaved to receive social support from strangers, fellow sufferers and other loved ones at any time and place rather than having to rely on someone in close physical proximity [5].

### 2.3 Managing Digital Assets

Users commonly struggle to manage their data due to: 1) large amounts of data they possess; 2) lack of motivation; and 3) the time and effort needed [12]. Despite the challenges, individuals find value in their digital assets, offering a sense of pride and fulfilment, making the management of their digital legacy worthwhile due to the significant value inherited digital assets can hold [12, 47]. There is a general societal concern for the management of digital legacies [10].  In a study done by [26] a little more than half (56%) of the respondents were worried about the management of their personal technologies after their death.

### 2.4 Perceptions

**Awareness about Digital Legacy.**

There is a lack of awareness regarding the management of digital assets and the potential to leave behind resource-intensive digital waste after one's death [29, 31]. [31] found that many online users have not considered their digital legacy but believed it should be considered in the future. Even among people who have a legal will, 70% had no clear understanding of what would happen to their digital assets after their passing [13]. [8] found that many students had never considered their own digital legacy unless they had experienced the passing of loved ones with active SNS.

**Attitudes towards Digital Legacy.**

Western society has traditionally viewed death as taboo. Although recent developments in technology have shifted people's attitudes towards death, individuals still tend to avoid making end-of-life decisions [28]. This can be explained by terror management theory that suggests that individuals avoid these decisions due to their belief that they are not going to die soon. This may explain the low usage of digital asset management tools. Additionally, people are generally not enthusiastic about planning for their death as it can involve tedious planning and thoughtful evaluation of what to leave behind [12]. People are motivated to plan their digital legacy if they view digital data as a gift as opposed to a burden, as it can then be framed as a meaningful process [12].

To safeguard one's digital assets, it is essential that people are educated about proper planning [8].

**Preferences towards Digital Legacy.**

There are varying preferences regarding digital legacy and how digital remains are managed after one's death. [25] researched how young people, who represent the internet generation, comprehend death and how that shapes their digital posthumous interaction. They found that 53.8% of the participants wished to leave a posthumous message that will be displayed after their death [25]. Similarly, the respondents in [44] study expressed a desire towards ensuring their valued digital artefacts are preserved, not only for themselves but for their loved ones. In contrast, [8] found that the majority of participants desired that their online digital remains are deleted, with only 24.4% wanting their own profiles to be active after their passing. Although the majority of these remains are text-based content and photographs at present, over time it is likely to expand to various other types of content as more applications and services are moved to the cloud.

The control and management of digital assets after death is another area of concern for individuals. Some people prefer to control which of their digital assets are to be kept or deleted, while others prefer automated alternatives [43]. A study found that 45%-50% of individuals preferred that someone is granted access to their personal email, social platforms, and digital accounts after their passing [18]. 31%-36% wished that all access be denied to their digital assets, and the remaining individuals preferred partial access [18].

**Religious beliefs.**

Cultural and religious beliefs influence how people view [25]. This can provide a guideline for SNS and digital legacy management platforms to respect users' digital legacy preferences and beliefs. Some research has been done on how rituals and practices can be enhanced through blending physical and digital interaction [34]. Religious farewell rituals demonstrate individuals' desire to maintain a connection with deceased loved ones [25]. Death is a social and cultural construct with specific sets of values and meanings. These cultural beliefs, rooted in religion, determine appropriate interactions with the dead [25]. [25] found two opposing ideas about death. Some believe that there is an abstract life beyond death while others view death as the end.

### 2.5    Challenges in Post-Mortem Data

Inheriting digital assets is complex due to the lack of established social, cultural, and religious guidelines [13, 34].  There are four main challenges associated with digital legacies after someone dies: 1) Problems with the transfer and access to assets due to authentication; 2) email access and password protection; 3) how to ensure the longevity of digital assets, and 4) concerns with the level of understanding of digital legacy terminology [13]. There is a trade-off between a service provider having access to one's data while providing privacy in exchange; however, once a user passes, the provider continues to have access [18]. This raises questions about post-mortem privacy rights, which is a person's right to control their digital legacy and assets after death [18].

### Policies And Regulations Issues.

In Common Law jurisdictions, privacy rights end when a person dies, but in the digital age online service providers continue to store and control the data, highlighting one of the challenges with post-mortem data  [28]. When someone dies, friends and family need the permission of service providers to gain access to data. This has resulted in legal complications and the few cases that have gone to court resulted in mixed outcomes [28]. Service providers often indicate in their terms and conditions that they have no legal obligation to grant access to data when a user passes away [47].  This is because their revenue comes from active users, which makes the allocation of resources to manage inactive accounts redundant [47]. The shift towards digital transformation, which includes online distribution of personal and sensitive information might cause stress for those receiving a digital legacy and for the curators thereof [33]. This is due to the fact that existing practices have not been adapted to the digital domain, even though its significance continues to grow [33]. The deceased's privacy rights can raise legal concerns regarding the extent of access an heir can have to an account [47]. Legally, three parties are affected by the contents of digital legacies: 1) the deceased individual who agreed to the terms of service; 2) the services they signed up for that established the rules; and 3) the heirs who may seek access to the deceased's account [47]. If the digital executor is left with account details, they have the potential to act as the deceased user and invade their privacy. This raises the ethical question of whether heirs should or should not be given access to deceased accounts.

## 3 Conceptual Background

Fig.1 outlines the adapted conceptual model which includes constructs from protection motivation theory, technology acceptance model, task-technology fit and theory of planned behaviour [2, 23, 37]. PMT explores the motives for one's digital legacy protective behaviours. TAM addresses the acceptance of new technology, while the TTF model focuses on the fit between task characteristics and technology characteristics. Finally, TPB integrates attitudes, societal norms, and perceived controls to influence one's behavioural intentions.

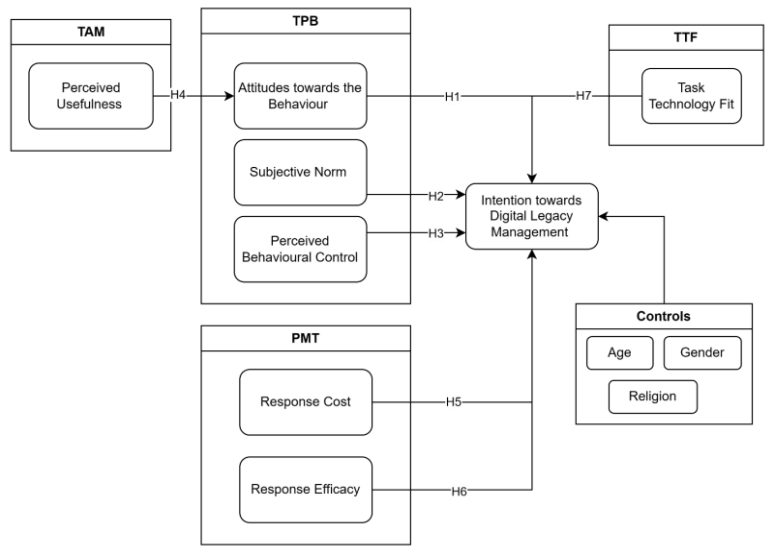

**Fig. 1.** Conceptual model adapted from [2, 23, 37].

### 3.1 Hypotheses

Hypotheses are important in deducing from theory and in subsequently, testing the hypotheses to come to a conclusion. In Table 1, the hypotheses developed for the study are presented.

**Table 1.** Hypotheses developed for the study.

| Hypotheses | Supporting Literature |
|---|---|
| 1. Attitudes towards digital legacy management outcomes have a positive influence on individuals' intentions to manage their digital legacy. | Attitudes and intentions can be influenced by ideas about the outcomes of a conduct .[2] |
| 2. Subjective norms of friends and family have a positive influence on digital legacy management intentions. | Subjective norm is a normative cognition that represents a person's assessment of whether significant people want them to engage in the target activity as well as their drive to comply with these others [15]. The stronger the positive perception of the subjective norm with respect to a behaviour, the more likely it is that there will be an intent to perform the behaviour [2]. |
| 3. Perceived behavioural control has a positive influence on digital legacy management intentions. | The combination of intentions and perceptions of behavioural control significantly contributes to the variation observed in behavioural intention [2]. Perceived behavioural control represents the capabilities and resources users possess, and lacking these essentials, their |

| Hypotheses | Supporting Literature |
| --- | --- |
| | intention towards a behaviour will be diminished [24]. |
| 4. Perceived usefulness of digital legacy management systems has a positive influence on attitudes towards digital legacy management intentions. | According to the Technology Acceptance Model (TAM), an individual's intention to accept technology is directly influenced by attitude which is influenced by perceived usefulness [1]. |
| 5. Response cost has a negative influence on digital legacy management intentions. | A higher perceived response cost will result in lower likelihood of individuals engaging in a specific protective behaviour [42]. |
| 6. Response efficacy has a positive influence on digital legacy management intentions. | Individuals who perceive a measure as effective are more likely to develop an intention to adopt it [19]. |
| 7. Task-technology fit has a positive influence on intention towards digital legacy management. | Studies suggest that task-technology fit leads to effective utilization. However, such utilization and enhanced performance cannot be achieved until the intention is realized [11]. |

# 4 Methodology

The study used a positive paradigm to employ scientific methods and adopted a deductive approach, employing Theory of Planned Behaviour, Technology Acceptance Model, Protection Motivation Theory and Task Technology Fit to formulate testable hypotheses. Utilizing a descriptive research design, the study examined the concepts and relationships related to digital legacy, a relatively new field, in order to establish a foundational understanding that can inform future studies.

## 4.1 Data Collection

A survey questionnaire was used to gather data from participants. This research instrument provided an adequate method of collecting data from a large sample while eliminating the need for the researcher to be present. This questionnaire was distributed to South African residents over the age of 18 using the platform Qualtrics. The questionnaire consisted of closed-ended structured questions that have been developed from previous studies, for simplicity, all questions were transformed to 5-item Likert with item responses ranging from 1 (strongly disagree) to 5 (strongly agree).

# 5 Data Analysis and Findings

The survey questionnaire was created, disseminated, and recorded using Qualtrics over a period of two weeks. Out of the initial 228 responses, 25 incomplete responses were removed, leaving a final sample of 203 valid responses. Data was analyses using PLS-SEM in SmartPLS.

## 5.1 Demographics Description

The demographics for age were segmented into five groups: 18-30, 31-40, 41-50, 51-60, and 61+. The majority of respondents (33.0%) were between the ages of 18-30 years old. 13.8% of respondents fell within the 31-40 age bracket. 12.3% of the participants were aged between 41-50 and 16.3% of respondents were between 51-60 years old. Lastly, 24.6% of respondents were 61 years or older.

## 5.2 Internal Consistency Reliability

Composite reliability was used to measure the internal consistency reliability, which is the preferred measure in PLS-SEM as it does not assume that each indicator is equally reliable [16, 46]. Composite reliability of values above 0.70 are satisfactory, as seen in Table 2, all constructs pass the composite reliability check.

**Table 2.** Composite Reliability.

| Construct | Composite Reliability |
|---|---|
| Attitudes | 0.920 |
| Intentions | 0.905 |
| Perceived Behavioural Control | 0.893 |
| Perceived Usefulness | 0.862 |
| Response Cost | 0.796 |
| Response Efficacy | 0.831 |
| Subjective Norms | 0.873 |
| Task Technology Fit | 0.933 |

### 5.3 Convergent Validity

Convergent Validity is tested using Average Variance Extracted (AVE) which indicates whether the latent construct can explain more than half of its indicators' variances [17]. An AVE of 0.50 or more indicates a sufficient degree of convergent reliability and will be accepted [16]. As seen in Table 3 the AVE extracted from the model shows that all constructs pass [36].

**Table 3.** Average Variance Extracted.

| Construct | Average Variance Extracted (AVE) |
|---|---|
| Attitudes | 0.743 |
| Intentions | 0.760 |
| Perceived Behavioural Control | 0.807 |
| Perceived Usefulness | 0.678 |
| Response Cost | 0.567 |
| Response Efficacy | 0.622 |
| Subjective Norms | 0.696 |
| Task Technology Fit | 0.823 |

### 5.4    Hypotheses Test Results

To assess the validity of the hypothesized relationships in the model, the path coefficients, t-values, and p-values are examined [4, 36]. Path coefficients, which typically fall between -1 and +1, signify the strength and direction of the relationship between constructs. Values closer to +1 suggest strong positive relationships, whereas values closer to -1 indicate strong negative relationships [36]. The p-value indicates the likelihood that a statistical outcome would occur due to chance [4]. A p-value lower than 0.05 is considered statistically significant at a 5% level, confirming the hypothesized relationship. T-values greater than 1.96 in two-tailed testing indicate a 5% level of statistical significance [16]. Table 4 shows each hypothesis path coefficients and its significance.

**Table 4.** Path coefficients of the structural model and significance testing results.

| Hy-pothesis | Path | Path Coeffi-cient | t-value | p-value | Sup-ported |
|---|---|---|---|---|---|
| H1 | ATT -> INT | 0.227 | 3.323 | 0.001 | **Yes** |
| H2 | SN -> INT | 0.285 | 3.850 | 0.000 | **Yes** |
| H3 | PBC -> INT | 0.345 | 5.658 | 0.000 | **Yes** |
| H4 | PU -> ATT | 0.352 | 4.919 | 0.000 | **Yes** |
| H5 | RC -> INT | -0.140 | 2.243 | 0.025 | **Yes** |
| H6 | RE -> INT | -0.006 | 0.095 | 0.924 | No |
| H7 | TTF -> INT | 0.045 | 0.611 | 0.541 | No |

## 6    Discussion

Based on the results, it was found that the attitude towards digital legacy management has a significant causal relationship on intention. This corroborates with the TPB framework by [24] that looks at users' behavioural intentions towards a digital communication tool. Furthermore, the results demonstrate that the perceived usefulness of managing one's digital legacy positively influences the attitude towards digital legacy management. This is in accordance with studies that explore individuals' acceptance of the usage of new technologies [1].

The findings indicate that subjective norms have a significant causal relationship with the intention to manage one's digital legacy. The results are similar with studies from [35, 38] that explore the digital management of banking. It highlights the importance of societal and peer perspectives in shaping an individual's intention towards digital legacy management. The results show that the demographic control gender has a significant influence on subjective norms. This could imply that societal norms affect genders differently with regards to digital legacy management. This is in line with gender studies that found peer influence to have a greater influence on women [27, 45].

The perceived behavioural control was also found to have a significant causal relationship on intention. This finding is consistent with a study done by [24] looking at users' behavioural intentions towards a digital communication tool. It implies that an individual's confidence in their capabilities and the resources play a role in influencing their intentions towards digital legacy management.

The results show that the response cost negatively influences digital legacy management intentions, which corroborates with previous literature looking at information security behaviours [21, 40, 42]. When individuals perceive a high cost associated with managing their digital legacy, they are less inclined to engage in digital legacy management. However, the results demonstrate that response efficacy's influence on intentions was insignificant and in the opposite direction than the hypothesised positive direction. This is contrary to numerous studies findings [14, 19, 40], although, there are instances where the anticipated positive influence of response efficacy on intentions were not supported [39, 41]. For individuals to actively manage and protect their digital assets and online presence after death, it is expected that they recognize the advantages of safeguarding these digital legacies [42]. In the demographic analysis, Figure 4 reveals that 82% of participants are either unfamiliar with digital legacy or are indifferent towards digital legacy management. This could be why response efficacy was not supported.

The findings revealed that the task-technology fit is insignificant and does not support intentions towards digital legacy management. This is contrary to previous studies looking at the perceived fit between technology and a task and the user's intention to use that technology [11, 20, 22]. This suggests that there is a poor alignment between individual needs in this domain and the technology solutions available. Another problem similar to response efficacy, is that participants are unfamiliar with the technologies designed for digital legacy management, preventing them from effectively understanding task-technology fit.

## 7 Conclusion

The rapid growth of digitalization in the past decades has highlighted the significance of managing one's digital legacy. This pertains to managing one's digital assets, including social media internet interactions and personal digital data. Understanding the intricacies of digital legacy management has become crucial for individuals, loved ones of the deceased and digital service providers. With South Africa's increasing digital integration into one's daily life, this study provides a unique context for understanding digital legacy management due to this country's digital divide and distinct digital legacy policies.

This study contributes to literature using the Theory of Planned Behaviour, Technology Acceptance Model, Protection Motivation Theory and Task-Technology Fit. By employing a multi-theoretical approach, this study has provided a broad perspective on what behavioural predictors influence individuals' intention to manage their digital legacy and how they do so in South Africa. The behavioural predictors are Attitudes, Subjective Norms, Perceived Behavioural Control, Perceived Usefulness and Response

Cost, which have all shown to be significant for digital legacy management intentions. These valuable insights hold significant potential for influencing platform design, awareness campaigns, and shaping policy frameworks tailored to a South African context. This research serves as a foundation for academicians, digital platforms, and technology companies in the digital legacy management field. Lastly, this paper emphasizes the need for platform developers to reassess and standardize post-mortem data designs and policies according to user preferences, resources and local regulations.